\documentclass[prd,twocolumn,amsmath,amssymb, superscriptaddress,floatfix]{revtex4}
\usepackage{bm}
\usepackage{amsmath}
\usepackage{epsf}
\usepackage{color}
\usepackage{natbib}
\usepackage{graphicx}
\usepackage{multirow}

\definecolor{darkgreen}{cmyk}{0.85,0.2,1.00,0.2}

\def\simlt{\lesssim}
\def\simgt{\gtrsim}

\begin{document}
\title{Non-linear evolution of $f(R)$ cosmologies II: power spectrum}

\author{Hiroaki Oyaizu}
\affiliation{Kavli Institute for Cosmological Physics,
and Enrico Fermi Institute,
University of Chicago, Chicago IL 60637}
\affiliation{Department of Astronomy \& Astrophysics, University of Chicago, Chicago IL 60637}
\author{Marcos Lima}
\affiliation{Kavli Institute for Cosmological Physics,
and Enrico Fermi Institute,
University of Chicago, Chicago IL 60637}
\affiliation{Department of Physics, University of Chicago, Chicago IL 60637}
\author{Wayne Hu}
\affiliation{Kavli Institute for Cosmological Physics,
and Enrico Fermi Institute,
University of Chicago, Chicago IL 60637}
\affiliation{Department of Astronomy \& Astrophysics, University of Chicago, Chicago IL 60637}

\begin{abstract}
We carry out a suite of cosmological simulations of modified action $f(R)$ models where cosmic
acceleration arises from an alteration of gravity instead of dark energy.  These models introduce an extra scalar
degree of freedom which enhances the force of gravity below the inverse mass or Compton scale of the scalar.  The simulations exhibit the so-called
chameleon mechanism, necessary for satisfying local constraints on gravity,
 where this scale depends on environment,
in particular the depth of the local gravitational potential.  We find that the chameleon
mechanism can substantially suppress the enhancement of power spectrum in the non-linear regime if the background field value is comparable to or smaller than the depth of the gravitational
potentials of typical structures.   Nonetheless power spectrum enhancements at 
intermediate scales remain at a measurable level for models even when the expansion history 
is indistinguishable from a cosmological constant, cold dark matter model.  Simple scaling
relations that take the linear power spectrum into a non-linear spectrum fail to capture
the modifications of $f(R)$ due to the change in collapsed structures, the chameleon 
mechanism, and the time evolution of the modifications.   \end{abstract}        
\date{\today}
\maketitle

\section{Introduction} \label{sec:intro}

Cosmic acceleration can arise from either an exotic form of energy with negative
pressure or a modification to gravity in the infrared.  
Self-consistent models for the latter
are highly constrained by the stringent tests of gravity in the solar system.
Additional propagating degrees of freedom must be suppressed by
non-linearities in their equations of motion
\cite{Vainshtein:1972sx,Deffayet:2001uk,Dvali:2006su} in a stable manner \cite{Dolgov:2003px,Sawicki:2007tf,Seifert:2007fr}.  These suppression mechanisms manifest 
themselves  with the formation of non-linear structure in the Universe
\cite{Lue:2004rj,Koyama:2007ih,hu07b}.
  Understanding the
physical content, phenomenology and even the basic viability of such models thus requires cosmological simulations.

One possibility that has received much recent attention is the so-called $f(R)$ class of models
(see \cite{Sotiriou:2008rp} and references therein).  
These models generate acceleration through a replacement
of the Einstein-Hilbert action by a function of
the Ricci or curvature scalar $R$  \cite{Caretal03,NojOdi03,Capozziello:2003tk}.  
They also introduce an extra
propagating scalar degree of freedom that acts as an effective fifth force on all forms of matter 
\cite{Chi03,Chiba:2006jp}.    
The range of the force depends non-linearly on the local curvature and can be made to become infinitesimal at high curvature.  With an appropriate choice of the function $f(R)$,
deep potential regions can trap the field at high curvature leading to a nonlinear
``chameleon mechanism'' \cite{khoury04a} that suppresses local deviations from 
ordinary gravity \cite{Navarro:2006mw,Faulkner:2006ub,hu07a}.  

Whether or not solar system tests of gravity are satisfied in an $f(R)$ model then depends
on the depth of the gravitational potential including the astrophysical and cosmological structure
surrounding it \cite{hu07a}.  Likewise observable deviations from ordinary gravity for
upcoming dark energy probes such as weak lensing, galaxy clustering and clusters
of galaxies depend on the whole history of non-linear structure
formation.

Previous cosmological simulations ({\it e.g.}~\cite{WhiKoc01,Maccio:2003yk,Nusser:2004qu,StaJai06,Laszlo:2007td}) have focused on modifications of the force law with a fixed and density independent range.
Such modifications alone are incapable of satisfying local tests of gravity. 

 In a companion 
paper \cite{oyaizu08b}, the numerical methodology for solving the non-linear field
equation of $f(R)$ gravity was established.  
In this second paper of the series,
we apply this methodology and carry out a suite of cosmological simulations of $f(R)$ models that are chosen 
to expose the impact of the chameleon mechanism on 
the power spectrum of the matter and the lensing potential.  

We begin in \S \ref{sec:fr} with a review of non-linear gravitational dynamics in $f(R)$ models
and proceed to the simulation results in \S \ref{sec:results}.  We discuss these results in
\S \ref{sec:conclusion}.

\section{$f(R)$ dynamics} \label{sec:fr}

\subsection{Basic equations}

The class of modifications we consider generalizes the
Einstein-Hilbert action to include an arbitrary function $f(R)$ of the scalar curvature $R$
 \begin{eqnarray}
S =  \int{d^4 x \sqrt{-g} \left[ \frac{R+f(R)}{16\pi G} + L_{m} \right]}\,. 
\label{eqn:action}
\end{eqnarray}
Here $L_{m}$ is the Lagrangian of the ordinary 
matter which remains minimally coupled.
Setting $f(R)=0$ recovers general relativity (GR) without a cosmological constant whereas 
setting $f(R) = -16\pi G \rho_\Lambda$ = const. recovers it with a cosmological constant.
Here and throughout $c=\hbar=1$.  

Variation of Eq.~(\ref{eqn:action}) with respect to the metric yields the modified Einstein equations
\begin{eqnarray}\label{eqn:metricvar}
G_{\alpha\beta} + F_{\alpha\beta}
&=&  8\pi G T_{\alpha\beta}\,,\\
F_{\alpha\beta} &=& f_{R} R_{\alpha\beta}-\left({f\over2} -\Box f_{R}\right) g_{\alpha\beta}
- \nabla_{\alpha}\nabla_{\beta}f_{R} \,, \nonumber
\end{eqnarray}
where the field,
\begin{equation}
f_R \equiv {d f(R) \over d R} ,
\end{equation}
plays the role of a propagating extra scalar degree of freedom.
In particular the trace of the modified Einstein equation (\ref{eqn:metricvar}) yields the
equation of motion for the field
\begin{equation}
\Box f_R =  {\partial V_{\rm eff} \over \partial f_R}\,,
\label{eqn:field}
\end{equation}
with the effective potential defined by
\begin{equation}
{\partial V_{\rm eff} \over \partial f_R} \equiv {1\over 3}\left[ R- f_R R + 2 f - 8\pi G(\rho-3p) \right]\,. \label{eqn:potn}
\end{equation}
The effective potential has an extremum at
 \begin{equation}
 R - R f_R + 2 f = 8\pi G (\rho-3p) \,,
 \end{equation}
and its 
curvature  is given by
 \begin{equation}
\mu^{2} = {\partial^2 V_{\rm eff} \over \partial f_R^2} = { 1\over 3} \left( {{1 + f_R \over df_{R}/dR} - R} \right)\,.
\label{eqn:mass}
 \end{equation}
This can be interpreted as the effective mass of the field $f_{R}$ and defines the
range of the force.  For stability, the extremum should be a minimum and
hence $\mu^2>0$ \cite{Sawicki:2007tf}.

Phenomenologically viable models typically must have very flat $f(R)$ functions
such that $|f_{R}|\ll 1$ at cosmological
curvatures and larger.  The model we simulate 
is in the class of $f(R)$ functions proposed in \cite{hu07a},
\begin{eqnarray}
f(R) \propto  {R \over A R + 1}   \,,
\label{eqn:fRfull}
\end{eqnarray}
where $A$ is a constant with dimensions of length squared.   In the
limit $R \rightarrow 0$, $f(R)
\rightarrow 0$ as with GR with no cosmological constant.
 For sufficiently high curvature that $AR \gg 1$, $f(R)$ can be approximated
as a constant, which drives the acceleration, plus a term that is inversely proportional to 
curvature.  In this limit, we can approximate Eq.~(\ref{eqn:fRfull}) by
\begin{eqnarray}
f(R) \approx -16 \pi G \rho_\Lambda - f_{R0} { \bar R_0^2 \over R} \,,
\end{eqnarray}
where we have set the constant $A$ to match some effective cosmological constant
$\rho_{\Lambda}$.   Here we define $\bar R_{0}=\bar R(z=0)$ as the background curvature today
and $f_{R0}= f_{R}(\bar R_{0})$.

Taking $|f_{R0}|\ll 1$, the background expansion follows the $\Lambda$CDM history with
the same $\rho_{\Lambda}$ to leading order in $|f_{R0}|$ \cite{hu07a}.  In particular the background curvature 
 may be approximated
as
\begin{equation}
\bar R \approx 3 H_{0}^{2} \left[\Omega_{\rm m} (1+z)^{3} + 4\Omega_\Lambda \right] \,,
\end{equation}
where $\Omega_i = 8\pi G\rho_i(z=0)/3H_0^2$.
We can also simplify the mass term in Eq.~(\ref{eqn:mass})
$\mu \approx (3 df_{R}/dR)^{-1/2}$ defining the comoving Compton wavelength or
range of
the field $\lambda_{C}$ as
\begin{eqnarray}
{\lambda_C \over 1+z} = \mu^{-1} \approx \sqrt{ 6 |f_{R0}| { R_{0}^{2} \over R^{3}}} \,.
\label{eqn:compton}
\end{eqnarray}
Notice that the range of the interaction has a steep inverse dependence on the local curvature $R$.

We take the
WMAP3 \citep{spergel07a} flat background cosmology throughout: $\Omega_{\rm m} = 0.24$, $\Omega_{\Lambda}
= 0.76$, $\Omega_{\rm b} = 0.04181$, $H_{0} = 73$ km/s/Mpc, and initial power
in curvature fluctuations  $A_{s}= (4.52 \times 10^{-5})^2$ at $k=0.05$Mpc$^{-1}$
with a tilt of $n_{s} = 0.958$.   
For $\Lambda$CDM these parameters give $\sigma_{8} = 0.76$.

For these values,  the Compton wavelength for the background today is
\begin{equation}
\lambda_{C0} \approx 3.2  \sqrt{ |f_{R0}|\over 10^{-6} } {\rm Mpc} \,.
\label{eqn:backgroundcompton}
\end{equation}
Note that even for field amplitudes as low as $|f_{R0}| \approx 10^{-7}$ where deviations from
$\Lambda$CDM
in the expansion history  are comparably negligible, the modifications to gravity at the
cosmological background density are order unity at astrophysically interesting scales of
a Mpc \cite{hu07a}.

Furthermore, the field equation (\ref{eqn:field}) can be simplified by neglecting
the small $f_R R$ term and assuming that matter dominates over radiation, thus resulting in
\begin{eqnarray}
\Box  f_{R}   = {1\over 3} \left[ R -8\pi G (\rho_{\rm m}+4 \rho_{\Lambda}) \right].
\end{eqnarray}
Subtracting off the background values for the field, curvature and density yields
\begin{eqnarray}
\Box \delta f_{R}   = {1\over 3}\left[  \delta R(f_{R})-8\pi G \delta \rho_{\rm m} \right],
\label{eqn:reducedfieldeqn}
\end{eqnarray}
where $\delta R \equiv R - \bar{R}$, 
$\delta f_{R} = f_R - f_{R}(\bar{R})$, $\delta \rho_{\rm m} = \rho_{\rm m} - \bar{\rho}_{\rm m}$.
Note that the field fluctuation is defined by subtracting off the field evaluated at the
background curvature and not the spatially averaged field value.  The procedure
eliminates the potential ambiguity of defining fluctuations with a highly non-linear
field equation.

The minimum of the effective potential is at a curvature corresponding to the GR
expectation $\delta R = 8\pi G \delta \rho_{\rm m}$.  If the field value achieves this minimum
then the high density regions have high curvature and short Compton wavelengths which 
suppress the deviations from ordinary gravity.
 However, we shall see in \S \ref{sec:chameleon}  that
whether the field value achieves this minimum at any given location depends on the
depth of the gravitational potential.

Finally, we work on scales much less than the horizon such that the quasi-static limit applies
where time derivatives may be neglected compared with spatial derivatives.  This
corresponds to assuming that the field instantaneously relaxes to its equilibrium
value.  More specifically,  relaxation must occur on a time scale that is short compared with the non-relativistic motion of particles.
This approximation should be excellent for wavelengths that are not 
orders of magnitude larger
than the Compton scale below which field perturbations propagate near the speed of light.
The field equation in comoving coordinates then becomes
 a non-linear Poisson-type equation
\begin{eqnarray}
\nabla^2 \delta  f_{R} = \frac{a^{2}}{3}\left[\delta R(f_{R}) - 8 \pi G \delta \rho_{\rm m}\right]. \label{eqn:frorig}
\end{eqnarray}
An explicit consistency test of this approximation in the cosmological context is given in
\cite{oyaizu08b}.

Since $f(R)$ is a metric theory of gravity, particles move in the metric or gravitational
potential in the same way as in general relativity.  However the field  acts as a source
that distinguishes the two potentials in the metric
\begin{equation}
ds^2 = -(1+2\Psi) dt^2 + a^2(1+2\Phi) dx^2 .
\label{eqn:lineelement}
\end{equation}
In the quasi-static limit, the modified Einstein equations (\ref{eqn:metricvar}) imply that the Newtonian potential $\Psi$, whose gradient is responsible
for the motion of particles, is given by a Poisson equation that is linear in $\delta \rho_{\rm m}$
and $\delta R$ \cite{Zha05},
\begin{eqnarray}
\nabla^2 \Psi = \frac{16 \pi G}{3}a^{2} \delta \rho_{\rm m} - \frac{a^{2}}{6} \delta R(f_R) \,.\label{eqn:potorig}
\end{eqnarray}
Equations (\ref{eqn:frorig}) and (\ref{eqn:potorig}) define a closed system 
for the gravitational potential given the density field.
It is interesting to note that braneworld modified gravity models \cite{Dvali:2000hr}
 obey a similar system
of equations except that the effective potential involves field gradients 
\cite{Koyama:2007ih}.

\subsection{Non-linear chameleon} \label{sec:chameleon}

Before proceeding to the numerical solution of these equations, it is worthwhile to examine the
qualitative aspects of the $f(R)$ system of equations to expose potential observational 
consequences.  In particular, these equations exhibit the chameleon mechanism under which
modifications to gravity become environment dependent.

 The force law modifications are manifested by the appearance of the second potential $\Phi$ in 
Eq.~(\ref{eqn:lineelement}).
Gravitational lensing and redshifts of photons depend on a combination of the two potentials
\begin{eqnarray}
\nabla^2 {(\Phi-\Psi) \over 2} =  -{4\pi Ga^2 } \delta \rho_{\rm m} \,,
\label{eqn:lenspot}
\end{eqnarray} 
Note that this relationship between the lensing potential and the matter density
is unaltered from the GR expectation.   It is the Poisson equation for $\Psi$ that
has an altered relationship to the matter density and governs the motion of non-relativistic
particles. 
Ordinary gravity is recovered if  $\delta R = 8 \pi G \delta \rho_{\rm m}$, and the 
equation for the gravitational potential reduces to the unmodified equation,
\begin{eqnarray}
\nabla^2 \Psi = 4\pi G a^{2} \delta \rho_{\rm m}\,,
\end{eqnarray}
and 
\begin{equation}
{(\Phi-\Psi) \over 2} = \Phi = -\Psi .
\end{equation}

Deviation from this relation locally are constrained by solar system tests of gravity 
at the level of $|(\Phi+\Psi)/\Psi| \lesssim 10^{-5}$ \cite{Will:2005va}.
These deviations are related to changes in the field through
the field equation (\ref{eqn:frorig}),
the Poisson equation (\ref{eqn:potorig}) and the lensing potential (\ref{eqn:lenspot})
\begin{eqnarray}
\nabla^{2}(\Phi+\Psi) = -\nabla^{2}\delta f_{R}\,.
\end{eqnarray}
If the force modifications are small, then $|(\Phi+\Psi)/\Psi| \ll 1$ for local contributions to the
potential.  

This relationship gives a rule of thumb for the appearance of the chameleon effect.  
Consider an isolated spherically symmetric structure embedded in the background density.
At the center of the object, the change in the field is related to the total depth of the
potentials as
\begin{eqnarray}
\Delta f_{R} = -(\Phi+\Psi) \,,
\end{eqnarray}
as long as both remain finite at the center.  

To drive the range of the force modification
to a scale much smaller than the background value, the field amplitude $|f_{R}|$ must be
substantially below its background value $|f_{\bar R}|$ at the center.  
For example, at the current epoch this 
requires
\begin{equation}
|\Delta f_{R}| \sim |f_{R0}| \sim |\Phi+\Psi| \ll |\Psi| \,.  
\end{equation}
This is a necessary condition for the appearance of a chameleon.
A chameleon also does not appear if most of the mass contributing to $\Psi$ is on scales
above the Compton scale of the background.  In this case $|\Phi+\Psi|$ remains much smaller
than $\Psi$ in this regime and the change in potential does not contribute to a substantial change in the field $\Delta f_R$.
Thus the smooth potential contributed by very large scale structure does not enter into the
chameleon suppression.

 If the local potential is not sufficiently deep, the field prefers to remain
smooth at the background value 
and not track $\delta R(f_{R})=8\pi G \delta \rho_{\rm m}$.  Energetically, the cost of high field gradients
prevents the field from lying at the local minimum of the effective potential
\cite{hu07a}.  
In this case the field remains near its background value and $\delta R(f_{R}) \ll 8\pi G\delta \rho_{\rm m}$.  Eq.~(\ref{eqn:potorig}) then implies that the strength of gravity is
a factor of $4/3$ greater than ordinary gravity.  

This consequence can alternatively be seen directly through Eq.~(\ref{eqn:frorig}).  If the field fluctuation is
small then $\delta R \approx (dR/df_R)\delta f_R$ and the field equation can be solved in
Fourier space as a linear Poisson equation
\begin{equation}
\delta f_R({\bf k}) = {1\over 3} {8\pi G a^2 \delta \rho_{\rm m}({\bf k})  \over k^2 + a^2\bar \mu^2}\,,
\end{equation}
where $\bar\mu=\mu(\bar R)$ and defines the Compton scale in the background through
Eq.~(\ref{eqn:compton}).
This solution combined with the Poisson equation (\ref{eqn:potorig}) gives
\begin{equation}
k^2 \Psi({\bf k})  = - 4\pi G \left( {4 \over 3} - {1 \over 3} {\mu^2 a^2 \over k^2 +\bar \mu^2 a^2}\right)
a^2 \delta \rho_{\rm m}({\bf k}) \, , \label{eqn:linearfr}
\end{equation}
which has an effective $G$ that depends on wavelength such that $G \rightarrow {4/3}G$ below
the Compton wavelength in the background.  Note that linearity in the density field is {\it not} 
assumed.   Hence we will use this field linearization approximation 
to test the effects of the chameleon in the simulations.

\begin{figure}
  \resizebox{84mm}{!}{\includegraphics[angle=0]{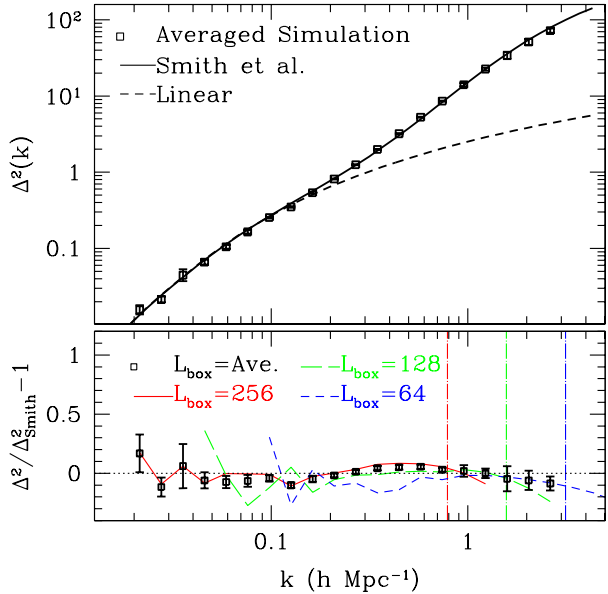}}
  \caption{\footnotesize The mean power spectra $\Delta^{2}=k^{3}P(k)/2\pi^{2}$ 
  of the cosmological simulations without 
  $f_R$ modifications ($\Lambda$CDM or $|f_{R0}|=0$).   {\it Upper panel}: Average power spectrum  of all simulations in comparison to linear theory and the Smith {\it et al.}~non-linear
  fit (solid line). 
  {\it Lower panel:}\ Relative power spectrum offset from Smith {\it et al.}~fit.  
  The individual simulation box sizes averages are plotted in solid, long-dashed, and dashed
  lines.   
   Results converge
  at approximately the half-Nyquist wavenumbers (vertical lines). 
  Points with errors show
      the mean and the 1-$\sigma$  error bars computed using the bootstrap method, weighted
  by simulation box volume,
 out to the half-Nyquist wavenumbers  (see text).  }
  \label{plot:pspec}
\end{figure}

In summary order unity modifications
to gravity are expected on scales smaller than the Compton scale of the background but
away from the centers of deep gravitational potential wells of cosmological structure.  
In these regions, the chameleon mechanism suppresses the local Compton scale and hence the
force modifications.


\section{$f(R)$ simulations} \label{sec:results}

\begin{figure}
  \resizebox{84mm}{!}{\includegraphics[angle=0]{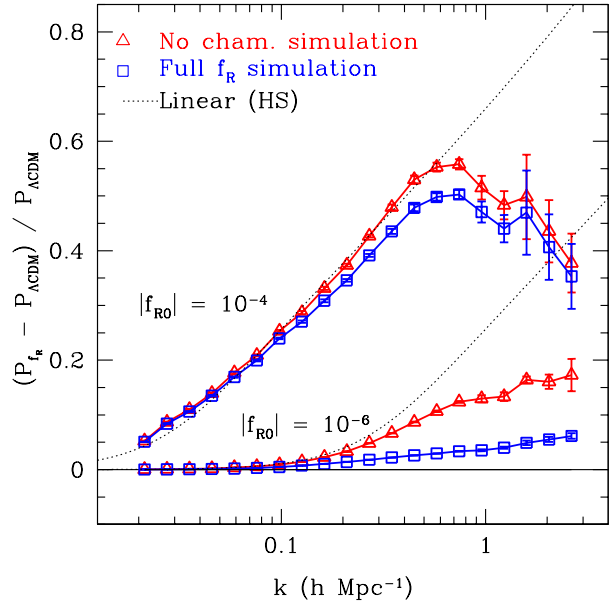}}
  \caption{\footnotesize Relative power spectrum enhancement over $\Lambda$CDM at $a=1$ for the full $f_R$ simulation compared with
  the no-chameleon $f_R$ simulations and linear theory.  At high $k$, linear theory underestimates the absolute power for both
  $\Lambda$CDM and $f_R$ while
  overestimating the relative enhancement.
  Without the chameleon, power is sharply enhanced
  on scales smaller than the Compton scale in the background which increases with $|f_{R0}|$
  (see Eq.~(\ref{eqn:backgroundcompton})).  For $|f_{R0}|=10^{-6}$ the chameleon strongly
  suppresses these enhancements at high $k$.  For $|f_{R0}|=10^{-4}$, this suppression
  is nearly absent except for a residual effect from the 
  chameleon at high redshift.    The 3 highest points in $k$ have increased sampling errors since
  they utilize only the $64 h^{-1}$ Mpc boxes.
  Note that the linear prediction is the fractional enhancement in the {\it linear} power spectrum,
  that is, $(P_{f_R,{\rm linear}} - P_{\rm {\Lambda}CDM, linear}) / P_{\rm {\Lambda}CDM, linear}$.
  }
  \label{plot:nocham}
\end{figure}

\begin{figure*}
  \resizebox{168mm}{!}{\includegraphics[angle=0]{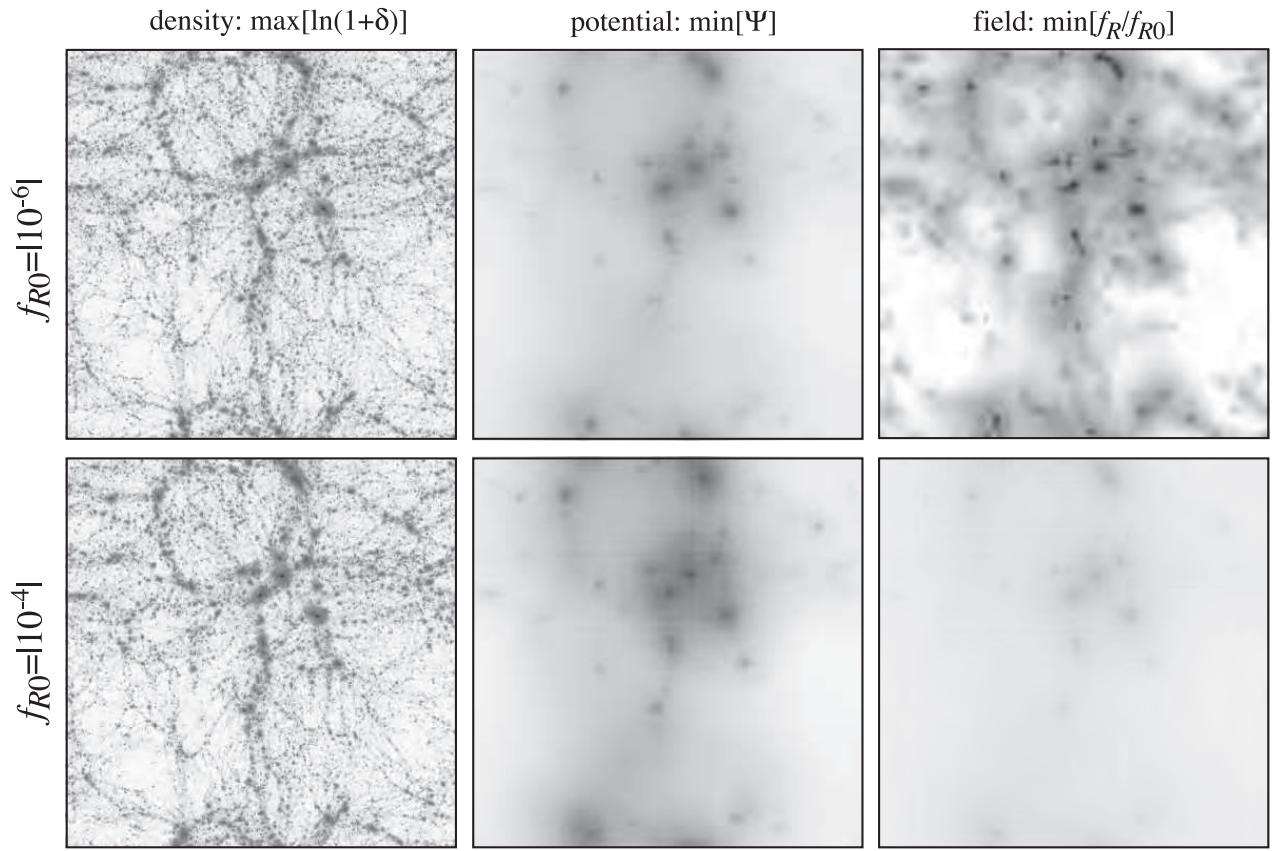}}
\caption{\footnotesize
2D slices through simulations at $a=1$  for 
various fields: matter overdensity $\delta \equiv \delta \rho_{\rm m}/\bar \rho_{\rm m}$ ({\it left});
 minimum gravitational potential $\Psi$ along the line of sight ({\it middle}) and minimum field $f_R/f_{R0}$
along the line of sight ({\it right}).
Each slice has the two dimensions of full $64 h^{-1}$~Mpc box and is
projected across 
$16 h^{-1}$~Mpc along the line of sight.
Shading scale ranges from white to black between the extreme
field values as follows:  
(-5.0 $\rightarrow$ 9.8) for $\ln(1+\delta)$; 
(0.3 $\rightarrow$ -4.3) x $10^{-5}$  for $\Psi$;
(1.05 $\rightarrow$ 0)  for $f_R/f_{R0}$. 
In the ${f}_{R0}=|10^{-6}|$ ({\it top}) simulation, the chameleon mechanism
suppresses the field and hence the force deviations in the deep potential
wells surrounding large overdensities.
In the 
${f}_{R0}=|10^{-4}|$ ({\it bottom}) simulation, the $f_R$ remains stiff
and a chameleon does not appear at the present.  While the density and potential
are slightly enhanced in this run due to the force modification, the appearance
of the chameleon depends mainly on the background field value $f_{R0}$.}
\label{plot:snapshots}
\end{figure*}

\subsection{Simulation description}

To solve the system of equations defined by the modified Poisson equation
(\ref{eqn:potorig}) and the quasi-static $f_R$ field equation (\ref{eqn:frorig}) in the context
of cosmological structure formation, 
we employ the methodology described in \cite{oyaizu08b}.
Briefly, the field equation for
 $f_R$ is solved on a regular
grid using relaxation techniques and multigrid iteration \citep{brandt73,briggs00a}.
The potential $\Psi$ is computed from the density and $f_R$ fields
using the fast Fourier transform method.
The dark matter particles are then moved according to the gradient of the
computed potential, $-\nabla \Psi$, using a second order accurate leap-frog integrator.

Since the Compton wavelength or range of the $f_R$ field in Eq.~(\ref{eqn:compton}) shrinks with increasing 
curvature, modifications to the force law and structure formation above
 any given comoving scale vanish at sufficiently high redshift.
 We can therefore treat the initial conditions for the simulations in the same way
 as in a cosmology with ordinary gravity. All simulations have the cosmological parameters given in the previous
section which make them all compatible with the high redshift CMB observations from WMAP
\cite{song07b}.
To explore the modifications induced by the $f_R$ field, we simulate models with
field strengths in the background of $|f_{R0} |=0$,  $10^{-6}$,   $10^{-5}$, $10^{-4}$.
Note that the $|f_{R0}|=0$ is exactly equivalent to $\Lambda$CDM.

Specifically, 
the initial conditions for the simulations are created using Enzo \citep{oshea04a},
a publicly available cosmological N-body + hydrodynamics code.
Enzo uses the Zel'dovich approximation to displace particles on a uniform grid according
to a given initial power spectrum.
We use the initial power spectra given by the transfer function of Eisenstein \& Hu \citep{eisenstein98a} and
normalization fixed at high $z$. 
Note that the initial spectrum does not include the effects of baryon acoustic oscillations.
The simulations are started at $z = 49$, and are integrated in time in
steps of $\Delta a = 0.002$.

In order to extend the dynamic range of the results, we run three simulations with box sizes $L_{\rm box}=256$ $h^{-1}$ Mpc,
128 $h^{-1}$ Mpc, and 64 $h^{-1}$ Mpc.
All simulations are run with 512 grid cells in each direction 
and
with $N_{\rm p}=256^3$ particles.  
Thus, the formal spatial resolutions of the simulations are 0.5 $h^{-1}$ Mpc, 0.25 $h^{-1}$
Mpc, and 0.125 $h^{-1}$ Mpc for the largest, middle, and smallest boxes, respectively.
The corresponding mass resolutions are $2.76\times 10^{11}h^{-1}\ \rm M_{\odot}$,
$3.45\times 10^{10}h^{-1}\ \rm M_{\odot}$, and $4.31\times 10^{9}h^{-1}\ \rm M_{\odot}$.

In the subsequent analysis, the particle Nyquist wave number of each simulation 
will play important roles in determining the range of trustworthy scales.
As shown in \cite{stabenau06a,Laszlo:2007td}, power spectrum of cosmological
simulations start to show systematic $>10\%$ deviations from Smith {\it et al.}~\cite{smith03a} 
and Peacock \& Dodds \cite{peacock96a} fits at wave numbers above half the particle
Nyquist mode, defined as $k_{N} = \pi N_{\rm p}^{1/3} / (2 L_{\rm box})$.
For our three simulation sizes, the half-Nyquist wave numbers are, in order of
decreasing box size, $0.79 h$ Mpc$^{-1}$, $1.57 h$ Mpc$^{-1}$, and $3.14 h$ Mpc$^{-1}$.

\begin{figure}
  \resizebox{84mm}{!}{\includegraphics[angle=0]{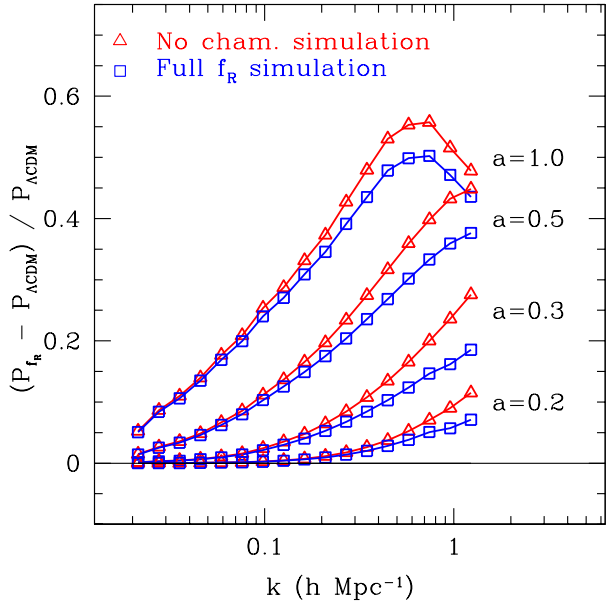}}
  \caption{\footnotesize Evolution of power spectrum deviation for $f_{R0} = |10^{-4}|$ for the full $f_{R}$
  simulation and the no-chameleon simulation.  The appearance of a chameleon
  at $a\simlt 0.5$ causes a large fractional change in the deviations at earlier epochs.  As the
  deviations grow, this offset remains as a smaller fraction of the total.
} 
  \label{plot:ctime2}
\end{figure}

Finally, to assess the impact of the chameleon mechanism on the power spectrum, we also carry out linearized $f_R$
simulations in which the gravitational potential, $\Psi$, is evaluated according to
Eq.~(\ref{eqn:linearfr}).
In the linearized treatment, the Compton wavelength is assumed to be
fixed by the background field and thus chameleon effects are not present.  Therefore, the difference between the full $f_R$ simulations and the linearized $f_R$
simulations are wholly due to the chameleon effects. To avoid
confusion with linearization of the density field, we will call these runs the 
``no-chameleon" simulations.

For each simulation box configuration, we run multiple simulations with different
realizations of the initial power spectrum in order to reduce finite sample variance.
The actual number of runs for each configurations are primarily constrained by
computational resources and are summarized in Table~\ref{table:runs}. 
To further reduce the sample variance, we average the difference of the statistics
per simulation from the $\Lambda$CDM run using the same realizations of the
initial conditions.

\begin{table}
\caption{Simulation type and number}
\begin{center}
  \leavevmode
  \begin{tabular}{c c c c c}
  && \multicolumn{3}{c}{$L_{\rm box}$ ($h^{-1}$ Mpc)} \vspace{1mm}\\
  \cline{3-5} 
  
&$|f_{R0}|$\ \  & $256 $\ \ \  & $128$\ \ \  & $64 $\ \ \  \\
\hline
\# of& $10^{-4}$ & 5 & 5 & 5\\
boxes&$10^{-5}$ & 5 & 5 & 5 \\
&$10^{-6}$ & 5 & 5 & 5 \\
&0 (GR) & 5 & 5 & 5 \\
\hline
\multicolumn{2}{c}{Spatial Resolution ($h^{-1}$ Mpc)} & 0.5 & 0.25 & 0.125 \\
\multicolumn{2}{c}{$k_{N}/2$ ($h$ Mpc$^{-1}$)} & 0.79 & 1.57 & 3.14 \\
\multicolumn{2}{c}{Mass Resolution  ($10^{10} h^{-1}\ \rm M_{\odot}$)}& $27.6$ & $3.45$ &  $0.431$ \\
\hline
\end{tabular}
\end{center}
\label{table:runs}
\end{table}

\subsection{Power spectrum results}

\begin{figure*}
  \resizebox{168mm}{!}{\includegraphics[angle=0]{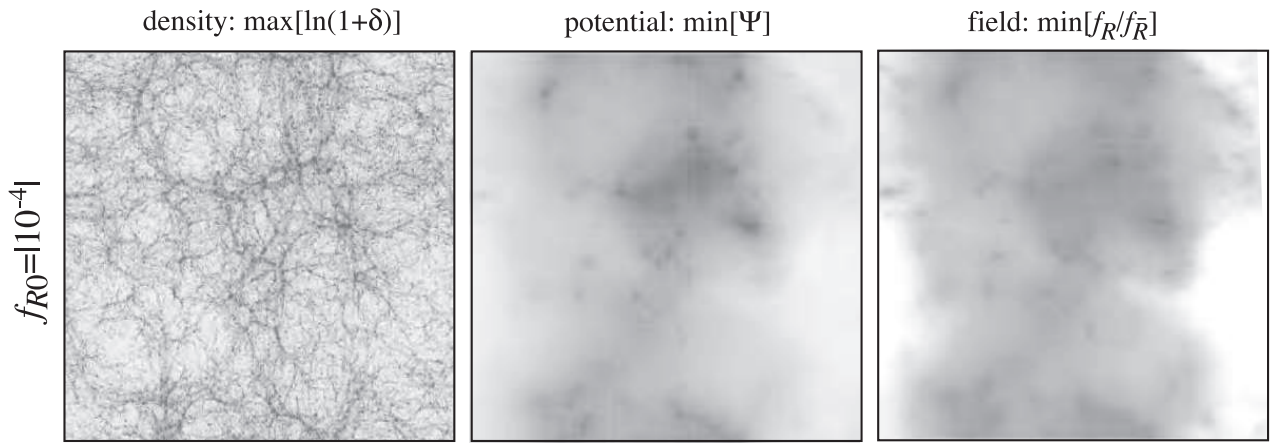}}
\caption{\footnotesize Same as in Fig.~\ref{plot:snapshots} but at $a=0.3$
for $|f_{R0}|=10^{-4}$ with the field values shown as a fraction of the background field
at that time $f_{\bar R}(a=0.3)$.  The field amplitude is suppressed in deep potential
wells exhibiting a chameleon that suppresses the early growth of structure.
The shading scale range for the various fields in this case is also identical 
to that of Fig.~\ref{plot:snapshots}. 
}
\label{plot:snapshotsz0.3}
\end{figure*}

We start with the pure $\Lambda$CDM  $f_{R0}=0$ simulations which serve as the baseline
reference for comparison with the other cases.
In Fig.~\ref{plot:pspec}, we show the average power spectrum in the simulations of the three 
box sizes.   The results from the various box sizes converge below about half the 
Nyquist wavenumber of the individual boxes.
For comparison we also show the fit of Smith {\it et al.}~\cite{smith03a}, 
which scales the linear theory predictions (also shown) into the non-linear regime.   The simulations converge
to the accuracy of the fit again at about half the Nyquist wavenumber.

All power spectra shown in this work, unless otherwise specified, are volume-weighted 
bootstrap-averaged from the multiple simulations of the same cosmological and $f(R)$ parameters.
In the averaging, the power spectra above the respective half-Nyquist modes are disregarded.
Similarly, the error bars represent the volume weighted bootstrap error, in which the individual
power spectrum data points are assumed to be uncorrelated.  We use 2000 bootstrap samples to compute the mean and the error.  When we show 
differences of power spectra, the differencing is performed before averaging.

In the bottom panel of Fig.~\ref{plot:pspec}, we show the relative difference between
our simulations and the Smith {\it et al.}~power spectra.
As expected, the mean simulation power spectrum matches the Smith {\it et al.}~results to $\sim 10\%$
for all scales of interest.
The error bars at the low $k$ end of the simulation spectrum are large due to the small 
number of largest scale modes in $L_{\rm box} = 256 h^{-1}$ Mpc box.
In the opposite end, the errors are larger due to large variation between the different
realizations of $L_{\rm box} = 64 h^{-1}$ Mpc boxes.   Note that bootstrap errors are
only a rough estimate of the true uncertainties in these regions where the sample
size is small.

Fig.~\ref{plot:nocham} shows the power spectrum enhancement of the $|f_{R0}| = 10^{-6}$ and $|f_{R0}| = 10^{-4}$ runs relative to the $\Lambda$CDM runs.   Note that
the enhancement of the power spectrum of the lensing potential $\Phi-\Psi$ is identical by
virtue of Eq.~(\ref{eqn:lenspot}).
For comparison we also
plot the linear theory predictions on the relative enhancement \cite{hu07a,pogosian08a} and the
no-chameleon $f_R$ simulation.    In both the linear theory and the no-chameleon results,
the force modification is completely determined by the background field.   Thus on
all scales smaller than the Compton wavelength in the background there is a sharp 
enhancement of power.  Linear theory tends to overestimate the enhancement due to
its neglect of mode coupling which makes the power at high $k$ dependent on scales
out to the non-linear scale where the effects are weaker.

Since non-linear structures can only make the Compton wavelength smaller, the no-chameleon and linear
predictions show the maximum scale out to which there are deviations from $\Lambda$CDM.
However in the full $f_R$ simulation the chameleon can dramatically change the results at 
high $k$ if the
background field amplitude is small enough to be overcome by the gravitational potentials
of collapsed objects as discussed in \S \ref{sec:fr}. 

 For the $|f_{R0}| = 10^{-6}$ case the enhancement is reduced
by a factor of 4 over the no-chameleon simulations at $k \approx 1 h$ Mpc$^{-1}$.
Previous simulations
of modified force laws have all been in models where there is no chameleon and no way to hide
deviations from ordinary gravity from local measurements \cite{WhiKoc01,Maccio:2003yk,Nusser:2004qu,StaJai06,Laszlo:2007td}.  
Note that even these reduced $\sim 1-10\%$ enhancements of power are potentially
observable in next generation weak lensing surveys ({\it e.g.}~\cite{Hu:2002rm,Knox:2006fh}).
As 
expected from the discussion in \S \ref{sec:fr}, for the $|f_{R0}|=10^{-4}$ model typical gravitational
potentials of order $10^{-6}-10^{-5}$ cannot overcome the background field and the
chameleon impact is greatly reduced.

\begin{figure}
  \resizebox{84mm}{!}{\includegraphics[angle=0]{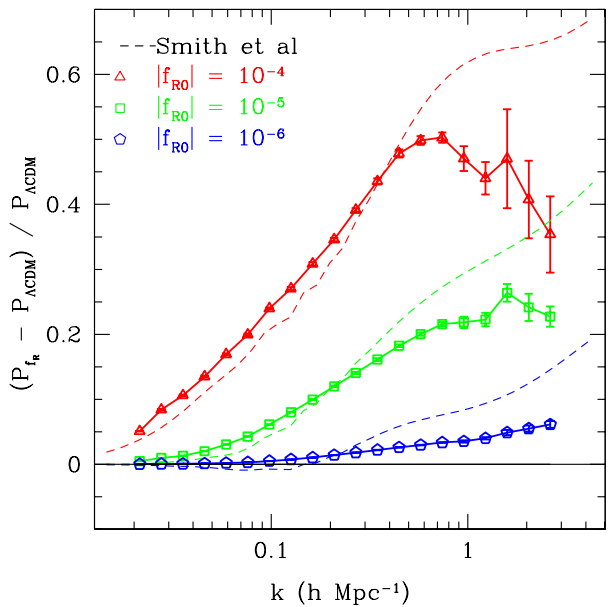}}
  \caption{\footnotesize Power spectrum deviations from $\Lambda$CDM of the full $f_R$ simulations vs.~the
  scaling predictions of Smith {\it et al.} employing linear $f_R$ calculations.  Note that in all cases the scaling predictions fail
  to capture the deviations at high $k$ due both to the change in the abundance and profiles
  of collapsed objects in the $f_R$ simulations and the chameleon mechanism.
 }
  \label{plot:psfrac.smith}
\end{figure}

By taking slices through the simulations we can see the effects of the environment 
dependence of the chameleon 
(see Fig.~\ref{plot:snapshots}).
In the $|f_{R0}|=10^{-6}$ run, the $f_R$ field is suppressed in the high density regions
which correspond to deep potential wells.   Notice that in low density regions the
force law is still modified and this accounts for the small enhancement of power over
the $\Lambda$CDM case that persists to high $k$ in Fig.~\ref{plot:nocham}. 
Thus two identical sets of objects separated by the same distance will
feel different forces depending on whether they are located in an overdense or underdense region.
Generic tests of gravity such as the comparison between dynamical and lensing mass
are predicted to produce null results in sufficiently overdense regions despite the
$\sim 1-10\%$ enhancement of power shown in Fig.~\ref{plot:nocham}.  Conversely,
even with a chameleon, substantial modifications to the gravitational force law can appear
in voids.

 In the $|f_{R0}|=10^{-4}$ run, the
field remains stiff and at its background value across the whole volume leading to
changes in the force law everywhere today.
Even in this run, the power spectrum is still suppressed compared
with the no-chameleon case (see Fig.~\ref{plot:nocham}).  This suppression comes about because the background
field value was substantially smaller at high redshift.  Structures that form during these
epochs are again affected by the chameleon and leave an impact at $z=0$ as
hierarchical structure formation progresses.   This can be seen in the evolution of the
power spectrum deviations in Fig.~\ref{plot:ctime2}. The impact of the chameleon at
high $z$ is fractionally large where the overall deviation is small.   The offset remains roughly constant at more recent 
epochs as the overall deviation increases.  In Fig.~\ref{plot:snapshotsz0.3} we show 
slices through the simulation at $a=0.3$ that reveal the presence of the chameleon in
deep potential wells at that time ({\it cf.} Fig.~\ref{plot:snapshots}).
The suppression of power spectrum enhancement at $k \gtrsim 0.7 h$ Mpc$^{-1}$ 
results from the shift of 1-halo contribution to larger scales and the relative flattening
of $\Delta^2(k)$, as shown in the halo model inspired treatment in \cite{hu07b}.

Finally, we can assess how well the Smith {\it et al.} \cite{smith03a} scaling works in the full $f_R$ 
simulations.  
In Fig.~\ref{plot:psfrac.smith} we show that this prescription fails to capture the
deviation from $\Lambda$CDM at high $k$.  This disagreement appears for all of the
$f_{R0}$ values and arises both from changes
in the contribution of collapsed objects to the power spectrum and the presence of the
chameleon effect.  

In fact, the whole concept of the linear power spectrum determining the
non-linear power spectrum at the same epoch that is shared by Smith {\it et al.}, the halo model, and 
other linear to non-linear
 scaling relations, is flawed in the context of a modification to gravity that evolves
with redshift.   We have seen that  the precise form  of  the $z=0$ power spectrum in the
$|f_{R0}|= 10^{-4}$ runs depends on the presence or absence of a chameleon at a higher
redshift.   This information is not directly encoded in the linear power spectrum.

\section{Discussion} \label{sec:conclusion}

We have carried out the first cosmological simulations of $f(R)$ models for cosmic
acceleration that exhibit the chameleon mechanism.  The chameleon mechanism
involves a non-linear field equation for the scalar degree of freedom that suppresses
the range of the gravitational force modification or Compton scale in the deep gravitational potential
wells of cosmological and astrophysical structure.  We have here focused on its impact
on the matter power spectrum or equivalently the potential power spectrum relevant
for weak lensing surveys.   While we have simulated only one  particular functional
form of $f(R)$, we expect that the qualitative behavior of other models that exhibit
a chameleon behavior  to yield similar results once scaled to the appropriate Compton
scales and field amplitudes.

 In the absence of
the chameleon mechanism, gravitational interactions would have an enhancement of
a factor of 4/3 on all scales smaller than the Compton scale in the cosmological background eventually leading to order unity enhancements in the
power at high wavenumber.  The chameleon mechanism turns on when the depth of the
local gravitational potential becomes comparable to the field amplitude in the background.  
We have shown through otherwise identical 
simulations of structure with the chameleon mechanism
artificially turned off that once the chameleon appears, it causes a substantial
reduction of the enhanced power.   For example, for a field amplitude of $|f_{R0}|=10^{-6}$
the change in the enhancement of power at $k \sim 1h^{-1}$Mpc is a factor of four.

Even in cases where current cosmological structures
do not possess a chameleon $(|f_{R0}| \simgt 10^{-5})$, there still is an impact on the
power spectrum due to evolutionary effects.  In $f(R)$ models where the field amplitude
decreases with curvature, the chameleon can appear again at high redshift when the
building blocks of current structure were assembled.

Scaling relations which take the linear power spectrum and map it into the non-linear
regime qualitatively misestimate the non-linear power spectrum in several ways.
For example, the Smith {\it et al.}~\cite{smith03a} prescription assumes that the non-linear power spectrum
depends only on the shape of the linear power spectrum near the non-linear scale.  This prescription
fails to describe both the chameleon mechanism and the change in the structure and
abundance of collapsed objects leading to a severe misestimate at high $k$.

A halo model can in principle do better to model these effects but simple prescriptions that scale the mass function to
the linear variance and leave halo profiles unchanged also do not describe the non-linear
effects to sufficient accuracy
({\it cf.}~\cite{hu07b}).  In the next paper of this series, we intend to study
 the impact of $f(R)$ modifications
 on halo properties.

\smallskip

\noindent {\it Acknowledgments}: We thank N. Dalal, B. Jain, J. Khoury,  
K. Koyama, A. Kravtsov, A. Upadhye, I. Sawicki, F. Schmidt, J. Tinker, and F. Stabenau
 for useful conversations.
This work was supported in part by the Kavli Institute for Cosmological
 Physics (KICP) at the University of Chicago through grants NSF PHY-0114422 and
 NSF PHY-0551142 and an endowment from the Kavli Foundation and its 
founder Fred Kavli.
HO was additionally supported by the NSF grants AST-0239759, AST-0507666, 
and AST-0708154 at the University of Chicago.
WH and ML  were additionally supported by
U.S.~Dept.\ of Energy contract DE-FG02-90ER-40560 and WH by 
 the David and Lucile Packard Foundation. 
Some of the computations used in this work have been performed on the Joint
Fermilab - KICP Supercomputing Cluster, supported by grants from Fermilab,
Kavli Insititute for Cosmological Physics, and the University of Chicago.

\bibliography{pspaper}

\end{document}